\documentclass[dvips]{actaprep}
\usepackage{supertabular,lscape,epsfig}
\usepackage{amssymb}
\usepackage{amsmath}
\DeclareSymbolFont{ppa}{OT1}{ppl}{m}{it}
\DeclareMathSymbol{\vv}{\mathalpha}{ppa}{'166}

\newfont{\hb}{rphvb at 10pt}
\newfont{\hbo}{rphvbo at 10pt}
\newfont{\bitt}{rptmbi at 12pt}
\newfont{\bits}{rptmbi at 11pt}

\renewcommand{\FigCap}[1]{\par\noindent Fig.\  %
  \refstepcounter{figure}\thefigure. #1\par}

\SetPages{83}{102}

\SetVol{61}{2011}

\begin{document}

\newcommand{\TabApp}[2]{\begin{center}\parbox[t]{#1}{\centerline{
  {\bf Appendix}}
  \vskip2mm
  \centerline{\small {\spaceskip 2pt plus 1pt minus 1pt T a b l e}
  \refstepcounter{table}\thetable}
  \vskip2mm
  \centerline{\footnotesize #2}}
  \vskip3mm
\end{center}}

\newcommand{\TabCapp}[2]{\begin{center}\parbox[t]{#1}{\centerline{
  \small {\spaceskip 2pt plus 1pt minus 1pt T a b l e}
  \refstepcounter{table}\thetable}
  \vskip2mm
  \centerline{\footnotesize #2}}
  \vskip3mm
\end{center}}

\newcommand{\TTabCap}[3]{\begin{center}\parbox[t]{#1}{\centerline{
  \small {\spaceskip 2pt plus 1pt minus 1pt T a b l e}
  \refstepcounter{table}\thetable}
  \vskip2mm
  \centerline{\footnotesize #2}
  \centerline{\footnotesize #3}}
  \vskip1mm
\end{center}}

\newcommand{\MakeTableApp}[4]{\begin{table}[p]\TabApp{#2}{#3}
  \begin{center} \TableFont \begin{tabular}{#1} #4 
  \end{tabular}\end{center}\end{table}}

\newcommand{\MakeTableSepp}[4]{\begin{table}[p]\TabCapp{#2}{#3}
  \begin{center} \TableFont \begin{tabular}{#1} #4 
  \end{tabular}\end{center}\end{table}}

\newcommand{\MakeTableee}[4]{\begin{table}[htb]\TabCapp{#2}{#3}
  \begin{center} \TableFont \begin{tabular}{#1} #4
  \end{tabular}\end{center}\end{table}}

\newcommand{\MakeTablee}[5]{\begin{table}[htb]\TTabCap{#2}{#3}{#4}
  \begin{center} \TableFont \begin{tabular}{#1} #5 
  \end{tabular}\end{center}\end{table}}


\newfont{\bb}{ptmbi8t at 12pt}
\newfont{\bbb}{cmbxti10}
\newfont{\bbbb}{cmbxti10 at 9pt}
\newcommand{\uprule}{\rule{0pt}{2.5ex}}
\newcommand{\douprule}{\rule[-2ex]{0pt}{4.5ex}}
\newcommand{\dorule}{\rule[-2ex]{0pt}{2ex}}
\def\thefootnote{\fnsymbol{footnote}}

\begin{Titlepage}
\Title{The Optical Gravitational Lensing Experiment.\\ OGLE-III Photometric
Maps of the Galactic Bulge Fields\footnote{Based on observations obtained
with the 1.3~m Warsaw telescope at the Las Campanas Observatory of the
Carnegie Institution for Science.}}
\Author{
M.\,K.~~S~z~y~m~a~ñ~s~k~i$^1$,~~
A.~~U~d~a~l~s~k~i$^1$,~~ 
I.~~S~o~s~z~y~ñ~s~k~i$^1$,~~
M.~~K~u~b~i~a~k$^1$,\\
G.~~P~i~e~t~r~z~y~ñ~s~k~i$^{1,2}$,~~ 
R.~~P~o~l~e~s~k~i$^1$,~~
£.~~W~y~r~z~y~k~o~w~s~k~i$^{3,1}$~~and~~
K.~~U~l~a~c~z~y~k$^1$}
{$^1$Warsaw University Observatory, Al.~Ujazdowskie~4, 00-478~Warszawa, Poland\\
e-mail: (msz,udalski,soszynsk,mk,pietrzyn,rpoleski,kulaczyk)@astrouw.edu.pl\\
$^2$Universidad de Concepción, Departamento de Astronomia,
Casilla 160--C, Concepción, Chile\\
$^3$Institute of Astronomy, University of Cambridge, Madingley Road,
Cambridge CB3~0HA,~UK\\
e-mail: wyrzykow@ast.cam.ac.uk}
\vspace*{-9pt}
\Received{June 15, 2011}
\end{Titlepage}

\Abstract{We present OGLE-III Photometric Maps of the Galactic bulge fields
observed during the third phase of the OGLE project. This paper describes
the last, concluding set of maps based on OGLE-III data.

The maps contain precise, calibrated {\it VI} photometry of about 340
million stars from 267 fields in the Galactic bulge observed during entire
OGLE-III phase (2002--2009), covering about 92 square degrees in the
sky. Precise astrometry of these objects is also provided.

We briefly discuss the photometry procedures and the quality of the
data. We also present sample data and color--magnitude diagrams of the
observed fields.

All photometric data are available to the astronomical community from the
OGLE Internet archive.}{Galaxy: bulge -- Surveys -- Catalogs -- Techniques:
photometric}

\Section{Introduction}
The Optical Gravitational Lensing Experiment (OGLE) is a long term project
(started in 1992) with scientific goals including gravitational
microlensing, search for extrasolar planets, variable stars etc. All these
goals require regular monitoring of millions of stars in dense stellar
fields located in the Galactic center, Magellanic Clouds and selected parts
of the Galactic disk. The photometry of all these stars collected during
the project lifetime constitutes a unique, huge data set that can be used
in many astrophysical applications.

Since the second phase of the project, OGLE-II, we have published precisely
calibrated, both photometrically and astrometrically, general {\it BVI} or
{\it VI} catalogs of objects called OGLE Photometric Maps. These Maps cover
the main project targets including the Magellanic Clouds, Galactic disk and
the Galactic center.

The OGLE maps were often used, for example, for modeling the structure of
the observed objects (\eg Rattenbury \etal 2007, Subramaniam and
Subramanian 2009, Nataf \etal 2010), determination of the interstellar
extinction (\eg Udalski \etal 2003b, Haschke, Grebel and Duffau 2011) or in
many analyzes of microlensing events (\eg Bennett \etal 2010, Abe \etal
2004). Also, the maps constitute a huge set of the secondary photometric
standards that can be used for calibration of photometry.

This paper presents the OGLE-III Photometric Map of the Galactic bulge,
concluding the most recent series of the OGLE Photometric Maps which
contain the photometry of objects observed during the third phase of the
project. The previous papers of this series have covered the stellar fields
in the Large and Small Magellanic Clouds (Udalski \etal 2008bc) and the
Galactic disk (Szymañski \etal 2010). The Galactic bulge photometry
constitutes the biggest data set in the series, containing almost 340
million stars observed in 267 dense fields of the center of Galaxy
covering 92 square degrees.

\Section{Observations}
Observations were carried out at Las Campanas Observatory in Chile,
operated by the Carnegie Institution for Science, using the 1.3-m Warsaw
Telescope equipped with the eight-chip mosaic CCD camera (Udalski 2003a)
covering $35\times35$~arcmin in the sky with the scale of
0.26~arcsec/pixel.

OGLE-III photometry of the Galactic center contains data obtained in the
{\it V}- and {\it I}-band filters which are close to the standard {\it VI}
bands. Only for very red objects, $(V-I)>2$~mag, the data obtained through
the OGLE-III glass {\it I} filter could not be precisely calibrated. This small
deficiency has been recently fixed using calibrations based on the first
OGLE-IV data, obtained with a new, 32-chip CCD camera equipped with
interferometric {\it I}-band filter which very accurately reproduces the
standard {\it I} pass-band. The details of this procedure will be given in
Section~3.

Observations of the Galactic bulge fields consist of 89\,175 exposures
spanning almost eight years between June 2001 and May 2009. The vast
majority of frames (87\,787) were obtained in {\it I}-band. The remaining
1\,388 frames were taken in {\it V} filter in order to collect some color
data for the observed stars. This asymmetric strategy proved to be the most
effective in terms of search for microlensing phenomena -- the main
scientific goal of the project -- and was adopted in all phases of the
experiment. The exposure times of the majority of frames were 120\,s for
{\it I}-band and 200\,s for {\it V}-band.

The data were taken only in good seeing and transparency conditions. The
seeing limit for Galactic bulge fields was set to $\approx1\zdot\arcs8$ but
it was not automatically enforced so the databases include some frames
taken at worse conditions. The median seeing of the {\it I}-band images is
equal to $1\zdot\arcs2$.

Table~1 lists all Galactic bulge fields covered by the OGLE-III Photometric
Maps giving their names, equatorial and galactic coordinates, number of
frames ({\it I} and {\it V}) included in the databases, the number of stars
detected in {\it I}-band and the mean value of $(V-I)$ color. Schematic map
of the fields on grids of equatorial and galactic coordinates is shown in
Fig.~1. The total area covered by the fields is about 92 square degrees (79
and 13 square degrees south and north of galactic equator,
respectively). About 7\% of this area, 6.3 square degrees, is covered by
the overlapping parts of adjacent fields. This includes both overlaps of
the neighboring main fields, resulting from the design of the Galactic
bulge pointings (Fig.~1) and overlaps of the subfields of adjacent chips of
the mosaic which are possible due to limited accuracy of telescope
pointing. We used this opportunity to construct reference images which are
bigger than single frames: $2180\times4176$ \vs $2048\times4096$ pixels per
chip (see Udalski \etal 2008a for details), thus covering, at least partly,
the gaps between CCD mosaic chips.

\renewcommand{\arraystretch}{1.05}
\MakeTableSepp{cccrrrrcc}{12.5cm}{OGLE-III fields in Galactic bulge}
{\hline 
\multicolumn{1}{c}{\douprule Field} & 
\multicolumn{1}{c}{RA} & 
\multicolumn{1}{c}{Dec (J2000)} &
\multicolumn{1}{c}{$l_{II}$} & 
\multicolumn{1}{c}{$b_{II}$} & 
\multicolumn{1}{c}{Nfr$_I$} & 
\multicolumn{1}{c}{Nfr$_V$} & 
\multicolumn{1}{c}{Nobj} & 
\multicolumn{1}{c}{$\langle V-I\rangle$} \\ 
\hline
\uprule
BLG100 & 17\uph51\upm00\zdot\ups0 & $-29\arcd59\arcm48\arcs$ & 359.6961 & $-1.5504$ & 2393 & 8 & 1484138 & 2.79 \\
BLG101 & 17\uph53\upm40\zdot\ups2 & $-29\arcd49\arcm52\arcs$ & 0.1331   & $-1.9643$ & 2418 & 34 & 2269274 & 2.10 \\
BLG102 & 17\uph56\upm20\zdot\ups2 & $-29\arcd30\arcm51\arcs$ & 0.6991   & $-2.3049$ & 2540 & 17 & 2026597 & 2.04 \\
BLG103 & 17\uph56\upm20\zdot\ups2 & $-30\arcd06\arcm22\arcs$ & 0.1865   & $-2.6018$ & 1393 & 4 & 1858589 & 2.07 \\
BLG104 & 17\uph59\upm00\zdot\ups3 & $-29\arcd28\arcm17\arcs$ & 1.0264   & $-2.7867$ & 1304 & 7 & 1917899 & 1.76 \\
BLG105 & 18\uph01\upm39\zdot\ups6 & $-29\arcd28\arcm12\arcs$ & 1.3138   & $-3.2882$ & 928 & 16 & 1950115 & 1.73 \\
BLG106 & 17\uph46\upm29\zdot\ups6 & $-37\arcd09\arcm46\arcs$ & 353.0553 & $-4.4247$ & 130 & 3 & 1160605 & 1.58 \\
BLG107 & 17\uph49\upm22\zdot\ups6 & $-37\arcd09\arcm45\arcs$ & 353.3508 & $-4.9179$ & 128 & 2 & 1100541 & 1.47 \\
BLG108 & 17\uph46\upm29\zdot\ups3 & $-36\arcd34\arcm09\arcs$ & 353.5649 & $-4.1178$ & 307 & 3 & 1330549 & 1.72 \\
BLG109 & 17\uph49\upm21\zdot\ups4 & $-36\arcd34\arcm11\arcs$ & 353.8606 & $-4.6125$ & 142 & 3 & 1156136 & 1.53 \\
BLG110 & 17\uph52\upm13\zdot\ups4 & $-36\arcd34\arcm11\arcs$ & 354.1533 & $-5.1087$ & 127 & 3 & 1306752 & 1.43 \\
BLG111 & 17\uph55\upm05\zdot\ups4 & $-36\arcd34\arcm11\arcs$ & 354.4427 & $-5.6070$ & 124 & 3 & 1090425 & 1.43 \\
BLG112 & 17\uph46\upm30\zdot\ups1 & $-35\arcd59\arcm04\arcs$ & 354.0683 & $-3.8184$ & 266 & 3 & 1158383 & 1.78 \\
BLG113 & 17\uph49\upm19\zdot\ups8 & $-35\arcd59\arcm04\arcs$ & 354.3628 & $-4.3094$ & 246 & 3 & 1281475 & 1.49 \\
BLG114 & 17\uph52\upm10\zdot\ups0 & $-35\arcd59\arcm04\arcs$ & 354.6549 & $-4.8038$ & 207 & 3 & 1155206 & 1.35 \\
BLG115 & 17\uph54\upm59\zdot\ups1 & $-35\arcd58\arcm46\arcs$ & 354.9463 & $-5.2946$ & 230 & 3 & 1204776 & 1.41 \\
BLG116 & 17\uph46\upm28\zdot\ups9 & $-35\arcd23\arcm15\arcs$ & 354.5784 & $-3.5065$ & 208 & 2 & 1188410 & 1.67 \\
BLG117 & 17\uph49\upm17\zdot\ups7 & $-35\arcd23\arcm16\arcs$ & 354.8735 & $-3.9985$ & 315 & 2 & 1270330 & 1.52 \\
BLG118 & 17\uph52\upm06\zdot\ups8 & $-35\arcd23\arcm17\arcs$ & 355.1659 & $-4.4934$ & 209 & 3 & 1335721 & 1.38 \\
BLG119 & 17\uph54\upm55\zdot\ups7 & $-35\arcd23\arcm17\arcs$ & 355.4551 & $-4.9895$ & 250 & 3 & 1232697 & 1.44 \\
BLG120 & 17\uph57\upm44\zdot\ups7 & $-35\arcd23\arcm17\arcs$ & 355.7413 & $-5.4878$ & 126 & 3 & 1245666 & 1.40 \\
BLG121 & 17\uph46\upm28\zdot\ups2 & $-34\arcd47\arcm43\arcs$ & 355.0849 & $-3.1982$ & 650 & 5 & 1736742 & 1.77 \\
BLG122 & 17\uph49\upm17\zdot\ups3 & $-34\arcd47\arcm50\arcs$ & 355.3814 & $-3.6954$ & 942 & 23 & 1722815 & 1.60 \\
BLG123 & 17\uph52\upm05\zdot\ups1 & $-34\arcd47\arcm52\arcs$ & 355.6737 & $-4.1900$ & 123 & 3 & 1588137 & 1.41 \\
BLG124 & 17\uph54\upm53\zdot\ups3 & $-34\arcd47\arcm52\arcs$ & 355.9641 & $-4.6874$ & 122 & 2 & 1429450 & 1.47 \\
BLG125 & 17\uph57\upm40\zdot\ups1 & $-34\arcd47\arcm45\arcs$ & 356.2507 & $-5.1815$ & 123 & 2 & 1408780 & 1.32 \\
BLG126 & 18\uph00\upm28\zdot\ups4 & $-34\arcd47\arcm45\arcs$ & 356.5351 & $-5.6830$ & 121 & 2 & 1289096 & 1.27 \\
BLG127 & 18\uph03\upm16\zdot\ups0 & $-34\arcd47\arcm45\arcs$ & 356.8153 & $-6.1841$ & 118 & 2 & 1073591 & 1.20 \\
BLG128 & 18\uph06\upm03\zdot\ups9 & $-34\arcd47\arcm45\arcs$ & 357.0931 & $-6.6880$ & 116 & 2 & 1020448 & 1.16 \\
BLG129 & 17\uph43\upm43\zdot\ups2 & $-34\arcd12\arcm11\arcs$ & 355.2961 & $-2.4063$ & 873 & 4 & 1518947 & 2.09 \\
BLG130 & 17\uph46\upm30\zdot\ups3 & $-34\arcd12\arcm12\arcs$ & 355.5959 & $-2.8981$ & 816 & 5 & 1580442 & 1.97 \\
BLG131 & 17\uph49\upm17\zdot\ups3 & $-34\arcd12\arcm12\arcs$ & 355.8926 & $-3.3915$ & 645 & 5 & 1502103 & 1.87 \\
BLG132 & 17\uph52\upm04\zdot\ups3 & $-34\arcd12\arcm14\arcs$ & 356.1858 & $-3.8870$ & 319 & 3 & 1693143 & 1.71 \\
BLG133 & 17\uph54\upm51\zdot\ups3 & $-34\arcd12\arcm14\arcs$ & 356.4764 & $-4.3842$ & 327 & 3 & 1714126 & 1.48 \\
BLG134 & 17\uph57\upm38\zdot\ups2 & $-34\arcd12\arcm14\arcs$ & 356.7638 & $-4.8829$ & 326 & 3 & 1702292 & 1.35 \\
BLG135 & 18\uph00\upm25\zdot\ups2 & $-34\arcd12\arcm14\arcs$ & 357.0484 & $-5.3838$ & 118 & 2 & 1335327 & 1.39 \\
BLG136 & 18\uph03\upm12\zdot\ups0 & $-34\arcd12\arcm14\arcs$ & 357.3296 & $-5.8858$ & 114 & 2 & 1176068 & 1.35 \\
BLG137 & 18\uph05\upm59\zdot\ups0 & $-34\arcd12\arcm14\arcs$ & 357.6083 & $-6.3902$ & 112 & 2 & 1089365 & 1.21 \\
BLG138 & 17\uph45\upm14\zdot\ups8 & $-33\arcd36\arcm41\arcs$ & 355.9667 & $-2.3677$ & 903 & 8 & 1410387 & 2.27 \\
BLG139 & 17\uph47\upm59\zdot\ups7 & $-33\arcd36\arcm42\arcs$ & 356.2631 & $-2.8574$ & 803 & 6 & 1411746 & 2.17 \\
BLG140 & 17\uph50\upm44\zdot\ups7 & $-33\arcd36\arcm52\arcs$ & 356.5544 & $-3.3505$ & 629 & 6 & 1349063 & 2.13 \\
BLG141 & 17\uph53\upm29\zdot\ups9 & $-33\arcd36\arcm47\arcs$ & 356.8467 & $-3.8439$ & 619 & 5 & 1503940 & 1.89 \\
BLG142 & 17\uph56\upm15\zdot\ups2 & $-33\arcd37\arcm03\arcs$ & 357.1312 & $-4.3424$ & 685 & 6 & 1597926 & 1.60 \\
BLG143 & 17\uph59\upm00\zdot\ups2 & $-33\arcd37\arcm03\arcs$ & 357.4160 & $-4.8396$ & 118 & 2 & 1366023 & 1.56 \\
\hline}

\setcounter{table}{0}
\MakeTableSepp{cccrrrrcc}{12.5cm}{Continued}
{\hline 
\multicolumn{1}{c}{\douprule Field} & 
\multicolumn{1}{c}{RA} & 
\multicolumn{1}{c}{Dec (J2000)} &
\multicolumn{1}{c}{$l_{II}$} & 
\multicolumn{1}{c}{$b_{II}$} & 
\multicolumn{1}{c}{Nfr$_I$} & 
\multicolumn{1}{c}{Nfr$_V$} & 
\multicolumn{1}{c}{Nobj} & 
\multicolumn{1}{c}{$\langle V-I\rangle$} \\ 
\hline
\uprule
BLG144 & 18\uph01\upm44\zdot\ups6 & $-33\arcd36\arcm46\arcs$ & 357.7011 & $-5.3344$ & 119 & 2 & 1261419 & 1.50 \\
BLG145 & 18\uph04\upm29\zdot\ups6 & $-33\arcd36\arcm45\arcs$ & 357.9805 & $-5.8349$ & 118 & 2 & 1176674 & 1.33 \\
BLG146 & 18\uph07\upm14\zdot\ups5 & $-33\arcd36\arcm46\arcs$ & 358.2563 & $-6.3371$ & 116 & 2 & 1056951 & 1.23 \\
BLG147 & 17\uph49\upm45\zdot\ups0 & $-33\arcd01\arcm31\arcs$ & 356.9554 & $-2.8709$ & 798 & 5 & 1608385 & 2.27 \\
BLG148 & 17\uph52\upm36\zdot\ups3 & $-33\arcd01\arcm32\arcs$ & 357.2603 & $-3.3861$ & 797 & 14 & 1660863 & 1.98 \\
BLG149 & 17\uph55\upm15\zdot\ups2 & $-33\arcd01\arcm33\arcs$ & 357.5402 & $-3.8657$ & 669 & 19 & 1301727 & 1.84 \\
BLG150 & 17\uph58\upm00\zdot\ups3 & $-33\arcd01\arcm33\arcs$ & 357.8285 & $-4.3657$ & 316 & 6 & 1485986 & 1.75 \\
BLG151 & 18\uph00\upm45\zdot\ups3 & $-33\arcd01\arcm31\arcs$ & 358.1142 & $-4.8669$ & 116 & 2 & 1381823 & 1.66 \\
BLG152 & 18\uph03\upm28\zdot\ups9 & $-33\arcd01\arcm14\arcs$ & 358.3982 & $-5.3635$ & 114 & 2 & 1322100 & 1.51 \\
BLG153 & 18\uph06\upm15\zdot\ups4 & $-33\arcd01\arcm31\arcs$ & 358.6761 & $-5.8753$ & 112 & 2 & 1202493 & 1.38 \\
BLG154 & 18\uph09\upm00\zdot\ups4 & $-33\arcd01\arcm31\arcs$ & 358.9527 & $-6.3818$ & 115 & 2 & 1040092 & 1.28 \\
BLG155 & 17\uph52\upm17\zdot\ups6 & $-32\arcd25\arcm59\arcs$ & 357.7386 & $-3.0292$ & 833 & 5 & 1776618 & 2.16 \\
BLG156 & 17\uph55\upm00\zdot\ups6 & $-32\arcd26\arcm00\arcs$ & 358.0281 & $-3.5242$ & 829 & 5 & 1831359 & 1.96 \\
BLG157 & 17\uph57\upm43\zdot\ups8 & $-32\arcd26\arcm00\arcs$ & 358.3155 & $-4.0214$ & 940 & 17 & 1596728 & 1.81 \\
BLG158 & 18\uph00\upm26\zdot\ups7 & $-32\arcd26\arcm02\arcs$ & 358.5989 & $-4.5197$ & 338 & 3 & 1468191 & 1.68 \\
BLG159 & 18\uph03\upm09\zdot\ups8 & $-32\arcd26\arcm01\arcs$ & 358.8807 & $-5.0198$ & 116 & 2 & 1277414 & 1.55 \\
BLG160 & 18\uph05\upm52\zdot\ups9 & $-32\arcd26\arcm01\arcs$ & 359.1595 & $-5.5218$ & 208 & 2 & 1144288 & 1.51 \\
BLG161 & 18\uph08\upm35\zdot\ups8 & $-32\arcd26\arcm00\arcs$ & 359.4354 & $-6.0246$ & 114 & 3 & 1018260 & 1.41 \\
BLG162 & 18\uph11\upm18\zdot\ups7 & $-32\arcd26\arcm00\arcs$ & 359.7083 & $-6.5291$ & 113 & 2 & 1007827 & 1.35 \\
BLG163 & 17\uph52\upm44\zdot\ups6 & $-31\arcd50\arcm13\arcs$ & 358.3014 & $-2.8090$ & 774 & 4 & 1471605 & 2.33 \\
BLG164 & 17\uph55\upm26\zdot\ups6 & $-31\arcd50\arcm12\arcs$ & 358.5912 & $-3.3041$ & 766 & 4 & 1752969 & 1.94 \\
BLG165 & 17\uph58\upm08\zdot\ups5 & $-31\arcd50\arcm13\arcs$ & 358.8776 & $-3.8008$ & 382 & 6 & 1693738 & 2.04 \\
BLG166 & 18\uph00\upm50\zdot\ups6 & $-31\arcd50\arcm14\arcs$ & 359.1616 & $-4.2998$ & 326 & 4 & 1785605 & 1.67 \\
BLG167 & 18\uph03\upm32\zdot\ups6 & $-31\arcd50\arcm15\arcs$ & 359.4427 & $-4.8001$ & 360 & 2 & 1502055 & 1.56 \\
BLG168 & 18\uph06\upm14\zdot\ups6 & $-31\arcd50\arcm17\arcs$ & 359.7208 & $-5.3022$ & 112 & 2 & 1393456 & 1.41 \\
BLG169 & 18\uph08\upm56\zdot\ups5 & $-31\arcd50\arcm19\arcs$ & 359.9961 & $-5.8055$ & 113 & 3 & 1106519 & 1.40 \\
BLG170 & 18\uph11\upm38\zdot\ups5 & $-31\arcd50\arcm18\arcs$ & 0.2696   & $-6.3104$ & 110 & 3 & 1129678 & 1.30 \\
BLG171 & 17\uph52\upm44\zdot\ups1 & $-31\arcd14\arcm47\arcs$ & 358.8100 & $-2.5080$ & 927 & 5 & 1562062 & 2.42 \\
BLG172 & 17\uph55\upm25\zdot\ups0 & $-31\arcd14\arcm47\arcs$ & 359.0997 & $-3.0028$ & 1406 & 6 & 1797708 & 1.97 \\
BLG173 & 17\uph58\upm06\zdot\ups0 & $-31\arcd14\arcm48\arcs$ & 359.3865 & $-3.4997$ & 786 & 5 & 1580697 & 2.03 \\
BLG174 & 18\uph00\upm46\zdot\ups9 & $-31\arcd14\arcm48\arcs$ & 359.6707 & $-3.9978$ & 344 & 2 & 1651630 & 1.85 \\
BLG175 & 18\uph03\upm27\zdot\ups8 & $-31\arcd14\arcm48\arcs$ & 359.9522 & $-4.4975$ & 384 & 3 & 1477697 & 1.72 \\
BLG176 & 18\uph06\upm08\zdot\ups9 & $-31\arcd14\arcm48\arcs$ & 0.2313   & $-4.9995$ & 355 & 2 & 1405589 & 1.56 \\
BLG177 & 18\uph08\upm49\zdot\ups7 & $-31\arcd14\arcm49\arcs$ & 0.5071   & $-5.5022$ & 115 & 3 & 1326372 & 1.39 \\
BLG178 & 18\uph11\upm32\zdot\ups2 & $-31\arcd14\arcm55\arcs$ & 0.7819   & $-6.0124$ & 114 & 3 & 1192805 & 1.33 \\
BLG179 & 17\uph50\upm00\zdot\ups0 & $-30\arcd39\arcm17\arcs$ & 359.0200 & $-1.7019$ & 1428 & 3 & 1110253 & 2.86 \\
BLG180 & 17\uph52\upm39\zdot\ups8 & $-30\arcd39\arcm20\arcs$ & 359.3117 & $-2.1949$ & 1463 & 9 & 1761449 & 2.41 \\
BLG181 & 17\uph55\upm19\zdot\ups7 & $-30\arcd39\arcm18\arcs$ & 359.6020 & $-2.6892$ & 1303 & 5 & 1727735 & 2.11 \\
BLG182 & 17\uph58\upm00\zdot\ups1 & $-30\arcd39\arcm17\arcs$ & 359.8903 & $-3.1869$ & 1433 & 8 & 1501630 & 2.12 \\
BLG183 & 18\uph00\upm39\zdot\ups6 & $-30\arcd39\arcm17\arcs$ & 0.1741   & $-3.6835$ & 803 & 5 & 1664547 & 1.82 \\
BLG184 & 18\uph03\upm18\zdot\ups0 & $-30\arcd39\arcm19\arcs$ & 0.4528   & $-4.1785$ & 768 & 5 & 1611072 & 1.77 \\
BLG185 & 18\uph06\upm00\zdot\ups6 & $-30\arcd39\arcm19\arcs$ & 0.7367   & $-4.6880$ & 821 & 4 & 1685043 & 1.64 \\
BLG186 & 18\uph08\upm39\zdot\ups7 & $-30\arcd39\arcm16\arcs$ & 1.0127   & $-5.1876$ & 111 & 4 & 1377094 & 1.45 \\
BLG187 & 18\uph11\upm19\zdot\ups8 & $-30\arcd39\arcm14\arcs$ & 1.2876   & $-5.6920$ & 107 & 2 & 1281754 & 1.27 \\
BLG188 & 17\uph59\upm00\zdot\ups2 & $-30\arcd03\arcm59\arcs$ & 0.5091   & $-3.0819$ & 814 & 8 & 1917577 & 1.81 \\
\hline}

\setcounter{table}{0}
\MakeTableSepp{cccrrrrcc}{12.5cm}{Continued}
{\hline 
\multicolumn{1}{c}{\douprule Field} & 
\multicolumn{1}{c}{RA} & 
\multicolumn{1}{c}{Dec (J2000)} &
\multicolumn{1}{c}{$l_{II}$} & 
\multicolumn{1}{c}{$b_{II}$} & 
\multicolumn{1}{c}{Nfr$_I$} & 
\multicolumn{1}{c}{Nfr$_V$} & 
\multicolumn{1}{c}{Nobj} & 
\multicolumn{1}{c}{$\langle V-I\rangle$} \\ 
\hline
\uprule
BLG189 & 18\uph01\upm39\zdot\ups1 & $-30\arcd03\arcm57\arcs$ & 0.7932   & $-3.5798$ & 776 & 12 & 1890245 & 1.59 \\
BLG190 & 18\uph04\upm28\zdot\ups1 & $-30\arcd03\arcm56\arcs$ & 1.0922   & $-4.1112$ & 1003 & 15 & 1685701 & 1.48 \\
BLG191 & 18\uph07\upm07\zdot\ups0 & $-30\arcd03\arcm56\arcs$ & 1.3705   & $-4.6125$ & 110 & 3 & 1537932 & 1.51 \\
BLG192 & 18\uph09\upm45\zdot\ups9 & $-30\arcd03\arcm56\arcs$ & 1.6463   & $-5.1153$ & 108 & 2 & 1416159 & 1.42 \\
BLG193 & 18\uph12\upm25\zdot\ups0 & $-30\arcd03\arcm57\arcs$ & 1.9196   & $-5.6204$ & 194 & 2 & 1368305 & 1.28 \\
BLG194 & 17\uph50\upm59\zdot\ups8 & $-29\arcd24\arcm27\arcs$ & 0.2022   & $-1.2486$ & 1379 & 11 & 1114100 & 2.62 \\
BLG195 & 17\uph53\upm38\zdot\ups4 & $-29\arcd14\arcm20\arcs$ & 0.6407   & $-1.6588$ & 1354 & 8 & 1937546 & 2.33 \\
BLG196 & 18\uph04\upm27\zdot\ups3 & $-29\arcd28\arcm06\arcs$ & 1.6137   & $-3.8178$ & 1298 & 14 & 1833526 & 1.54 \\
BLG197 & 18\uph07\upm04\zdot\ups3 & $-29\arcd28\arcm22\arcs$ & 1.8867   & $-4.3181$ & 915 & 6 & 1659192 & 1.52 \\
BLG198 & 18\uph09\upm42\zdot\ups3 & $-29\arcd28\arcm22\arcs$ & 2.1628   & $-4.8208$ & 766 & 5 & 1573823 & 1.40 \\
BLG199 & 18\uph12\upm20\zdot\ups3 & $-29\arcd28\arcm22\arcs$ & 2.4365   & $-5.3250$ & 114 & 3 & 1530714 & 1.28 \\
BLG200 & 17\uph26\upm59\zdot\ups7 & $-39\arcd54\arcm52\arcs$ & 348.6826 & $-2.6778$ & 117 & 2 & 887388 & 1.88 \\
BLG201 & 17\uph27\upm00\zdot\ups1 & $-39\arcd19\arcm30\arcs$ & 349.1736 & $-2.3510$ & 114 & 2 & 908803 & 2.13 \\
BLG202 & 17\uph29\upm58\zdot\ups6 & $-39\arcd54\arcm58\arcs$ & 348.9976 & $-3.1552$ & 111 & 2 & 831860 & 1.89 \\
BLG203 & 17\uph29\upm59\zdot\ups1 & $-39\arcd19\arcm30\arcs$ & 349.4930 & $-2.8317$ & 111 & 3 & 854256 & 2.05 \\
BLG204 & 18\uph14\upm59\zdot\ups9 & $-29\arcd28\arcm21\arcs$ & 2.7107   & $-5.8355$ & 110 & 3 & 1397880 & 1.18 \\
BLG205 & 17\uph57\upm16\zdot\ups4 & $-28\arcd53\arcm10\arcs$ & 1.3456   & $-2.1672$ & 1394 & 34 & 2275476 & 2.05 \\
BLG206 & 17\uph59\upm52\zdot\ups8 & $-28\arcd53\arcm00\arcs$ & 1.6324   & $-2.6606$ & 1363 & 18 & 2183459 & 1.62 \\
BLG207 & 18\uph02\upm33\zdot\ups4 & $-28\arcd52\arcm58\arcs$ & 1.9223   & $-3.1700$ & 1120 & 19 & 2002261 & 1.57 \\
BLG208 & 18\uph05\upm08\zdot\ups4 & $-28\arcd53\arcm00\arcs$ & 2.1987   & $-3.6636$ & 1399 & 41 & 1894677 & 1.52 \\
BLG209 & 18\uph07\upm43\zdot\ups9 & $-28\arcd53\arcm00\arcs$ & 2.4741   & $-4.1600$ & 108 & 4 & 1748973 & 1.50 \\
BLG210 & 18\uph10\upm22\zdot\ups1 & $-28\arcd53\arcm00\arcs$ & 2.7518   & $-4.6664$ & 23 & 2 & 1409499 & 1.43 \\
BLG211 & 18\uph12\upm59\zdot\ups1 & $-28\arcd53\arcm00\arcs$ & 3.0250   & $-5.1704$ & 22 & 2 & 1237744 & 1.30 \\
BLG212 & 18\uph15\upm36\zdot\ups6 & $-28\arcd52\arcm58\arcs$ & 3.2972   & $-5.6771$ & 102 & 3 & 1420104 & 1.25 \\
BLG213 & 18\uph18\upm13\zdot\ups1 & $-28\arcd53\arcm00\arcs$ & 3.5644   & $-6.1825$ & 20 & 4 & 1073979 & 1.25 \\
BLG214 & 17\uph57\upm32\zdot\ups6 & $-28\arcd17\arcm25\arcs$ & 1.8915   & $-1.9205$ & 1206 & 8 & 1862536 & 2.28 \\
BLG215 & 18\uph00\upm09\zdot\ups7 & $-28\arcd17\arcm26\arcs$ & 2.1784   & $-2.4206$ & 1082 & 8 & 1974275 & 1.86 \\
BLG216 & 18\uph02\upm44\zdot\ups5 & $-28\arcd17\arcm29\arcs$ & 2.4582   & $-2.9150$ & 1044 & 6 & 2010062 & 1.63 \\
BLG217 & 18\uph05\upm19\zdot\ups6 & $-28\arcd17\arcm27\arcs$ & 2.7373   & $-3.4111$ & 821 & 10 & 1921363 & 1.56 \\
BLG218 & 18\uph07\upm55\zdot\ups4 & $-28\arcd17\arcm28\arcs$ & 3.0146   & $-3.9113$ & 678 & 5 & 1713709 & 1.65 \\
BLG219 & 18\uph10\upm30\zdot\ups1 & $-28\arcd17\arcm32\arcs$ & 3.2868   & $-4.4098$ & 777 & 5 & 1647078 & 1.48 \\
BLG220 & 18\uph13\upm06\zdot\ups0 & $-28\arcd17\arcm32\arcs$ & 3.5598   & $-4.9129$ & 200 & 3 & 1717223 & 1.27 \\
BLG221 & 18\uph15\upm42\zdot\ups1 & $-28\arcd17\arcm32\arcs$ & 3.8308   & $-5.4180$ & 115 & 3 & 1557697 & 1.20 \\
BLG222 & 18\uph18\upm18\zdot\ups0 & $-28\arcd17\arcm30\arcs$ & 4.0997   & $-5.9236$ & 116 & 4 & 1347613 & 1.23 \\
BLG223 & 17\uph59\upm19\zdot\ups7 & $-27\arcd41\arcm56\arcs$ & 2.6011   & $-1.9671$ & 1323 & 30 & 1692197 & 2.28 \\
BLG224 & 18\uph01\upm55\zdot\ups2 & $-27\arcd41\arcm56\arcs$ & 2.8853   & $-2.4656$ & 1033 & 6 & 1916029 & 1.90 \\
BLG225 & 18\uph04\upm30\zdot\ups2 & $-27\arcd41\arcm58\arcs$ & 3.1657   & $-2.9641$ & 825 & 6 & 2159774 & 1.65 \\
BLG226 & 18\uph07\upm05\zdot\ups4 & $-27\arcd41\arcm59\arcs$ & 3.4443   & $-3.4645$ & 847 & 6 & 2078177 & 1.51 \\
BLG227 & 18\uph09\upm39\zdot\ups6 & $-27\arcd41\arcm59\arcs$ & 3.7191   & $-3.9629$ & 787 & 4 & 1986367 & 1.49 \\
BLG228 & 18\uph12\upm15\zdot\ups1 & $-27\arcd42\arcm00\arcs$ & 3.9937   & $-4.4670$ & 18 & 3 & 1314414 & 1.33 \\
BLG229 & 18\uph14\upm50\zdot\ups1 & $-27\arcd42\arcm00\arcs$ & 4.2653   & $-4.9707$ & 18 & 3 & 1159502 & 1.20 \\
BLG230 & 18\uph17\upm25\zdot\ups1 & $-27\arcd42\arcm00\arcs$ & 4.5347   & $-5.4757$ & 17 & 2 & 1179505 & 1.26 \\
BLG231 & 18\uph20\upm00\zdot\ups1 & $-27\arcd42\arcm00\arcs$ & 4.8019   & $-5.9820$ & 17 & 5 & 1052213 & 1.32 \\
BLG232 & 17\uph59\upm59\zdot\ups7 & $-27\arcd06\arcm25\arcs$ & 3.1886   & $-1.8015$ & 745 & 4 & 1557796 & 2.42 \\
BLG233 & 18\uph02\upm37\zdot\ups0 & $-27\arcd06\arcm24\arcs$ & 3.4774   & $-2.3085$ & 623 & 3 & 1739011 & 2.31 \\
\hline}

\setcounter{table}{0}
\MakeTableSepp{cccrrrrcc}{12.5cm}{Continued}
{\hline 
\multicolumn{1}{c}{\douprule Field} & 
\multicolumn{1}{c}{RA} & 
\multicolumn{1}{c}{Dec (J2000)} &
\multicolumn{1}{c}{$l_{II}$} & 
\multicolumn{1}{c}{$b_{II}$} & 
\multicolumn{1}{c}{Nfr$_I$} & 
\multicolumn{1}{c}{Nfr$_V$} & 
\multicolumn{1}{c}{Nobj} & 
\multicolumn{1}{c}{$\langle V-I\rangle$} \\ 
\hline
\uprule
BLG234 & 18\uph05\upm12\zdot\ups1 & $-27\arcd06\arcm25\arcs$ & 3.7592   & $-2.8102$ & 640 & 5 & 1981513 & 1.66 \\
BLG235 & 18\uph07\upm47\zdot\ups1 & $-27\arcd06\arcm26\arcs$ & 4.0386   & $-3.3128$ & 685 & 20 & 1823097 & 1.65 \\
BLG236 & 18\uph10\upm20\zdot\ups1 & $-27\arcd06\arcm28\arcs$ & 4.3118   & $-3.8105$ & 300 & 2 & 1772717 & 1.71 \\
BLG237 & 18\uph12\upm55\zdot\ups2 & $-27\arcd06\arcm29\arcs$ & 4.5867   & $-4.3162$ & 105 & 7 & 1642632 & 1.40 \\
BLG238 & 18\uph15\upm29\zdot\ups7 & $-27\arcd06\arcm29\arcs$ & 4.8586   & $-4.8211$ & 106 & 6 & 1553326 & 1.33 \\
BLG239 & 18\uph18\upm05\zdot\ups1 & $-27\arcd06\arcm30\arcs$ & 5.1296   & $-5.3303$ & 15 & 2 & 1093822 & 1.35 \\
BLG240 & 18\uph20\upm39\zdot\ups8 & $-27\arcd06\arcm30\arcs$ & 5.3975   & $-5.8384$ & 14 & 2 & 975542 & 1.39 \\
BLG241 & 18\uph06\upm02\zdot\ups2 & $-26\arcd30\arcm52\arcs$ & 4.3681   & $-2.6842$ & 790 & 13 & 1633314 & 1.97 \\
BLG242 & 18\uph08\upm36\zdot\ups3 & $-26\arcd30\arcm53\arcs$ & 4.6467   & $-3.1869$ & 564 & 5 & 1882684 & 1.84 \\
BLG243 & 18\uph11\upm10\zdot\ups3 & $-26\arcd30\arcm55\arcs$ & 4.9226   & $-3.6907$ & 187 & 3 & 1503175 & 1.66 \\
BLG244 & 18\uph13\upm44\zdot\ups4 & $-26\arcd30\arcm56\arcs$ & 5.1967   & $-4.1959$ & 171 & 4 & 1440633 & 1.52 \\
BLG245 & 18\uph16\upm18\zdot\ups5 & $-26\arcd30\arcm56\arcs$ & 5.4689   & $-4.7024$ & 15 & 3 & 1126469 & 1.44 \\
BLG246 & 18\uph18\upm49\zdot\ups0 & $-26\arcd30\arcm59\arcs$ & 5.7319   & $-5.1986$ & 104 & 3 & 1350870 & 1.42 \\
BLG247 & 18\uph21\upm24\zdot\ups1 & $-26\arcd31\arcm00\arcs$ & 6.0014   & $-5.7109$ & 14 & 2 & 853708 & 1.46 \\
BLG248 & 18\uph23\upm58\zdot\ups1 & $-26\arcd30\arcm59\arcs$ & 6.2673   & $-6.2205$ & 103 & 2 & 1092107 & 1.28 \\
BLG249 & 18\uph06\upm00\zdot\ups2 & $-25\arcd55\arcm18\arcs$ & 4.8827   & $-2.3890$ & 588 & 5 & 1477672 & 2.30 \\
BLG250 & 18\uph08\upm32\zdot\ups6 & $-25\arcd55\arcm26\arcs$ & 5.1581   & $-2.8895$ & 509 & 4 & 1723868 & 1.92 \\
BLG251 & 18\uph11\upm07\zdot\ups5 & $-25\arcd55\arcm27\arcs$ & 5.4376   & $-3.3985$ & 551 & 4 & 1553925 & 1.72 \\
BLG252 & 18\uph13\upm39\zdot\ups1 & $-25\arcd55\arcm20\arcs$ & 5.7108   & $-3.8968$ & 259 & 2 & 1405414 & 1.67 \\
BLG253 & 18\uph16\upm11\zdot\ups9 & $-25\arcd55\arcm21\arcs$ & 5.9822   & $-4.4015$ & 108 & 2 & 1459393 & 1.45 \\
BLG254 & 18\uph18\upm45\zdot\ups0 & $-25\arcd55\arcm21\arcs$ & 6.2522   & $-4.9082$ & 108 & 9 & 1342950 & 1.42 \\
BLG255 & 18\uph21\upm18\zdot\ups7 & $-25\arcd55\arcm30\arcs$ & 6.5190   & $-5.4192$ & 108 & 2 & 1158691 & 1.49 \\
BLG256 & 18\uph23\upm51\zdot\ups6 & $-25\arcd55\arcm31\arcs$ & 6.7843   & $-5.9277$ & 103 & 2 & 1101295 & 1.41 \\
BLG257 & 18\uph08\upm30\zdot\ups0 & $-25\arcd19\arcm48\arcs$ & 5.6739   & $-2.5939$ & 109 & 2 & 1485247 & 2.12 \\
BLG258 & 18\uph11\upm01\zdot\ups8 & $-25\arcd19\arcm49\arcs$ & 5.9493   & $-3.0950$ & 107 & 2 & 1487206 & 2.03 \\
BLG259 & 18\uph13\upm33\zdot\ups8 & $-25\arcd19\arcm49\arcs$ & 6.2232   & $-3.5979$ & 105 & 2 & 1331882 & 1.92 \\
BLG260 & 18\uph16\upm06\zdot\ups1 & $-25\arcd20\arcm00\arcs$ & 6.4929   & $-4.1045$ & 99 & 3 & 1499790 & 1.64 \\
BLG261 & 18\uph18\upm38\zdot\ups1 & $-25\arcd20\arcm00\arcs$ & 6.7627   & $-4.6098$ & 100 & 2 & 1368037 & 1.44 \\
BLG262 & 18\uph21\upm09\zdot\ups9 & $-25\arcd19\arcm59\arcs$ & 7.0303   & $-5.1154$ & 97 & 2 & 1218733 & 1.38 \\
BLG263 & 18\uph23\upm41\zdot\ups7 & $-25\arcd20\arcm00\arcs$ & 7.2955   & $-5.6225$ & 99 & 5 & 1171702 & 1.41 \\
BLG264 & 18\uph26\upm13\zdot\ups7 & $-25\arcd20\arcm00\arcs$ & 7.5593   & $-6.1312$ & 98 & 3 & 915042 & 1.40 \\
BLG265 & 18\uph14\upm29\zdot\ups4 & $-24\arcd44\arcm27\arcs$ & 6.8428   & $-3.5025$ & 106 & 4 & 1329570 & 1.86 \\
BLG266 & 18\uph17\upm01\zdot\ups4 & $-24\arcd44\arcm27\arcs$ & 7.1154   & $-4.0093$ & 102 & 5 & 1324518 & 1.72 \\
BLG267 & 18\uph19\upm34\zdot\ups1 & $-24\arcd44\arcm30\arcs$ & 7.3865   & $-4.5200$ & 21 & 4 & 886425 & 1.53 \\
BLG268 & 18\uph22\upm06\zdot\ups1 & $-24\arcd44\arcm30\arcs$ & 7.6551   & $-5.0291$ & 22 & 4 & 878867 & 1.46 \\
BLG269 & 18\uph24\upm38\zdot\ups1 & $-24\arcd44\arcm30\arcs$ & 7.9217   & $-5.5394$ & 23 & 4 & 973815 & 1.41 \\
BLG270 & 18\uph27\upm10\zdot\ups1 & $-24\arcd44\arcm30\arcs$ & 8.1864   & $-6.0507$ & 22 & 13 & 877532 & 1.43 \\
BLG271 & 18\uph29\upm42\zdot\ups1 & $-24\arcd44\arcm30\arcs$ & 8.4493   & $-6.5632$ & 20 & 3 & 675396 & 1.37 \\
BLG272 & 18\uph16\upm00\zdot\ups1 & $-24\arcd09\arcm00\arcs$ & 7.5274   & $-3.5254$ & 21 & 3 & 1039961 & 1.87 \\
BLG273 & 18\uph18\upm31\zdot\ups1 & $-24\arcd09\arcm00\arcs$ & 7.7985   & $-4.0318$ & 21 & 5 & 1096212 & 1.55 \\
BLG274 & 18\uph21\upm02\zdot\ups1 & $-24\arcd09\arcm00\arcs$ & 8.0677   & $-4.5393$ & 20 & 3 & 990320 & 1.49 \\
BLG275 & 18\uph23\upm33\zdot\ups1 & $-24\arcd09\arcm00\arcs$ & 8.3349   & $-5.0479$ & 20 & 5 & 847642 & 1.53 \\
BLG276 & 18\uph26\upm04\zdot\ups1 & $-24\arcd09\arcm00\arcs$ & 8.6003   & $-5.5575$ & 23 & 4 & 762870 & 1.52 \\
BLG277 & 18\uph28\upm35\zdot\ups1 & $-24\arcd09\arcm00\arcs$ & 8.8638   & $-6.0683$ & 22 & 5 & 733393 & 1.54 \\
BLG278 & 18\uph31\upm04\zdot\ups9 & $-24\arcd08\arcm58\arcs$ & 9.1240   & $-6.5758$ & 100 & 2 & 756314 & 1.49 \\
\hline}

\setcounter{table}{0}
\MakeTableSepp{cccrrrrcc}{12.5cm}{Continued}
{\hline 
\multicolumn{1}{c}{\douprule Field} & 
\multicolumn{1}{c}{RA} & 
\multicolumn{1}{c}{Dec (J2000)} &
\multicolumn{1}{c}{$l_{II}$} & 
\multicolumn{1}{c}{$b_{II}$} & 
\multicolumn{1}{c}{Nfr$_I$} & 
\multicolumn{1}{c}{Nfr$_V$} & 
\multicolumn{1}{c}{Nobj} & 
\multicolumn{1}{c}{$\langle V-I\rangle$} \\ 
\hline
\uprule
BLG279 & 18\uph16\upm00\zdot\ups1 & $-23\arcd33\arcm30\arcs$ & 8.0495   & $-3.2454$ & 20 & 4 & 1011305 & 1.81 \\
BLG280 & 18\uph18\upm30\zdot\ups1 & $-23\arcd33\arcm30\arcs$ & 8.3203   & $-3.7505$ & 19 & 4 & 989315 & 1.68 \\
BLG281 & 18\uph21\upm00\zdot\ups1 & $-23\arcd33\arcm30\arcs$ & 8.5891   & $-4.2568$ & 19 & 3 & 949372 & 1.54 \\
BLG282 & 18\uph23\upm30\zdot\ups1 & $-23\arcd33\arcm30\arcs$ & 8.8561   & $-4.7641$ & 17 & 3 & 800872 & 1.60 \\
BLG283 & 18\uph26\upm00\zdot\ups1 & $-23\arcd33\arcm30\arcs$ & 9.1213   & $-5.2725$ & 18 & 5 & 749195 & 1.51 \\
BLG284 & 18\uph28\upm30\zdot\ups1 & $-23\arcd33\arcm30\arcs$ & 9.3847   & $-5.7819$ & 18 & 3 & 657775 & 1.52 \\
BLG285 & 18\uph31\upm00\zdot\ups1 & $-23\arcd33\arcm30\arcs$ & 9.6463   & $-6.2923$ & 18 & 3 & 603916 & 1.47 \\
BLG286 & 18\uph16\upm00\zdot\ups1 & $-22\arcd58\arcm00\arcs$ & 8.5713   & $-2.9651$ & 17 & 4 & 915277 & 1.99 \\
BLG287 & 18\uph18\upm30\zdot\ups1 & $-22\arcd58\arcm00\arcs$ & 8.8435   & $-3.4723$ & 16 & 4 & 911160 & 1.81 \\
BLG288 & 18\uph21\upm00\zdot\ups1 & $-22\arcd58\arcm00\arcs$ & 9.1138   & $-3.9807$ & 16 & 4 & 881015 & 1.66 \\
BLG289 & 18\uph23\upm30\zdot\ups1 & $-22\arcd58\arcm00\arcs$ & 9.3822   & $-4.4901$ & 18 & 4 & 898691 & 1.57 \\
BLG290 & 18\uph26\upm00\zdot\ups1 & $-22\arcd58\arcm00\arcs$ & 9.6489   & $-5.0005$ & 14 & 3 & 788157 & 1.54 \\
BLG291 & 18\uph28\upm30\zdot\ups1 & $-22\arcd58\arcm00\arcs$ & 9.9138   & $-5.5120$ & 14 & 5 & 672798 & 1.49 \\
BLG292 & 18\uph31\upm00\zdot\ups1 & $-22\arcd58\arcm00\arcs$ & 10.1769  & $-6.0244$ & 15 & 5 & 614425 & 1.43 \\
BLG293 & 18\uph16\upm00\zdot\ups1 & $-22\arcd22\arcm30\arcs$ & 9.0929   & $-2.6845$ & 16 & 11 & 1040978 & 2.19 \\
BLG294 & 18\uph18\upm29\zdot\ups1 & $-22\arcd22\arcm30\arcs$ & 9.3646   & $-3.1905$ & 16 & 3 & 1021042 & 1.91 \\
BLG295 & 18\uph20\upm58\zdot\ups1 & $-22\arcd22\arcm30\arcs$ & 9.6345   & $-3.6974$ & 17 & 3 & 951091 & 1.78 \\
BLG296 & 18\uph23\upm27\zdot\ups1 & $-22\arcd22\arcm30\arcs$ & 9.9026   & $-4.2054$ & 15 & 5 & 936343 & 1.64 \\
BLG297 & 18\uph25\upm56\zdot\ups1 & $-22\arcd22\arcm30\arcs$ & 10.1689  & $-4.7144$ & 15 & 4 & 869260 & 1.45 \\
BLG298 & 18\uph28\upm25\zdot\ups1 & $-22\arcd22\arcm30\arcs$ & 10.4335  & $-5.2244$ & 13 & 3 & 838656 & 1.30 \\
BLG299 & 18\uph30\upm54\zdot\ups1 & $-22\arcd22\arcm30\arcs$ & 10.6964  & $-5.7354$ & 14 & 3 & 726365 & 1.34 \\
BLG300 & 18\uph17\upm00\zdot\ups1 & $-21\arcd47\arcm00\arcs$ & 9.7244   & $-2.6082$ & 17 & 3 & 1007743 & 2.00 \\
BLG301 & 18\uph19\upm28\zdot\ups1 & $-21\arcd47\arcm00\arcs$ & 9.9948   & $-3.1131$ & 17 & 3 & 892601 & 1.96 \\
BLG302 & 18\uph21\upm56\zdot\ups1 & $-21\arcd47\arcm00\arcs$ & 10.2635  & $-3.6190$ & 18 & 4 & 882631 & 1.95 \\
BLG303 & 18\uph24\upm24\zdot\ups1 & $-21\arcd47\arcm00\arcs$ & 10.5305  & $-4.1260$ & 17 & 5 & 934256 & 1.63 \\
BLG304 & 18\uph26\upm52\zdot\ups1 & $-21\arcd47\arcm00\arcs$ & 10.7957  & $-4.6339$ & 15 & 6 & 763531 & 1.56 \\
BLG305 & 18\uph29\upm19\zdot\ups8 & $-21\arcd47\arcm00\arcs$ & 11.0588  & $-5.1417$ & 16 & 5 & 694849 & 1.39 \\
BLG306 & 18\uph31\upm48\zdot\ups1 & $-21\arcd47\arcm00\arcs$ & 11.3213  & $-5.6525$ & 16 & 6 & 720132 & 1.42 \\
BLG307 & 18\uph23\upm10\zdot\ups1 & $-21\arcd11\arcm30\arcs$ & 10.9220  & $-3.5970$ & 18 & 3 & 780694 & 1.86 \\
BLG308 & 18\uph25\upm38\zdot\ups1 & $-21\arcd11\arcm30\arcs$ & 11.1894  & $-4.1063$ & 16 & 3 & 666080 & 1.75 \\
BLG309 & 18\uph28\upm06\zdot\ups1 & $-21\arcd11\arcm30\arcs$ & 11.4552  & $-4.6166$ & 16 & 4 & 655837 & 1.63 \\
BLG310 & 18\uph30\upm34\zdot\ups1 & $-21\arcd11\arcm30\arcs$ & 11.7194  & $-5.1278$ & 17 & 4 & 626459 & 1.64 \\
BLG311 & 18\uph33\upm02\zdot\ups1 & $-21\arcd11\arcm30\arcs$ & 11.9820  & $-5.6399$ & 17 & 4 & 568079 & 1.46 \\
BLG312 & 18\uph24\upm39\zdot\ups8 & $-20\arcd36\arcm00\arcs$ & 11.6095  & $-3.6311$ & 20 & 4 & 814800 & 1.89 \\
BLG313 & 18\uph27\upm07\zdot\ups1 & $-20\arcd36\arcm00\arcs$ & 11.8760  & $-4.1404$ & 19 & 4 & 682459 & 1.86 \\
BLG314 & 18\uph33\upm00\zdot\ups1 & $-20\arcd36\arcm00\arcs$ & 12.5082  & $-5.3646$ & 19 & 4 & 608451 & 1.49 \\
BLG315 & 17\uph52\upm16\zdot\ups1 & $-37\arcd10\arcm00\arcs$ & 353.6398 & $-5.4169$ & 19 & 2 & 797278 & 1.46 \\
BLG316 & 17\uph45\upm00\zdot\ups1 & $-37\arcd45\arcm30\arcs$ & 352.3905 & $-4.4789$ & 19 & 2 & 710766 & 1.80 \\
BLG317 & 17\uph47\upm54\zdot\ups1 & $-37\arcd45\arcm30\arcs$ & 352.6865 & $-4.9703$ & 18 & 2 & 734776 & 1.56 \\
BLG318 & 17\uph50\upm47\zdot\ups9 & $-37\arcd45\arcm21\arcs$ & 352.9810 & $-5.4620$ & 123 & 2 & 955689 & 1.47 \\
BLG319 & 17\uph47\upm30\zdot\ups1 & $-38\arcd21\arcm00\arcs$ & 352.1357 & $-5.2054$ & 19 & 2 & 756243 & 1.51 \\
BLG320 & 17\uph50\upm26\zdot\ups1 & $-38\arcd21\arcm00\arcs$ & 352.4295 & $-5.7005$ & 19 & 3 & 769101 & 1.40 \\
BLG321 & 17\uph53\upm22\zdot\ups1 & $-38\arcd21\arcm00\arcs$ & 352.7199 & $-6.1979$ & 20 & 3 & 636272 & 1.38 \\
BLG322 & 17\uph46\upm00\zdot\ups1 & $-38\arcd56\arcm30\arcs$ & 351.4748 & $-5.2575$ & 20 & 4 & 734816 & 1.54 \\
BLG323 & 17\uph48\upm57\zdot\ups1 & $-38\arcd56\arcm30\arcs$ & 351.7694 & $-5.7506$ & 20 & 3 & 706144 & 1.35 \\
\hline}
\renewcommand{\arraystretch}{1.1}

\setcounter{table}{0}
\MakeTableSepp{cccrrrrcc}{12.5cm}{Concluded}
{\hline 
\multicolumn{1}{c}{\douprule Field} & 
\multicolumn{1}{c}{RA} & 
\multicolumn{1}{c}{Dec (J2000)} &
\multicolumn{1}{c}{$l_{II}$} & 
\multicolumn{1}{c}{$b_{II}$} & 
\multicolumn{1}{c}{Nfr$_I$} & 
\multicolumn{1}{c}{Nfr$_V$} & 
\multicolumn{1}{c}{Nobj} & 
\multicolumn{1}{c}{$\langle V-I\rangle$} \\ 
\hline
\uprule
BLG324 & 17\uph51\upm54\zdot\ups1 & $-38\arcd56\arcm30\arcs$ & 352.0604 & $-6.2458$ & 19 & 3 & 603439 & 1.36 \\
BLG325 & 17\uph38\upm30\zdot\ups1 & $-39\arcd32\arcm00\arcs$ & 350.2073 & $-4.3283$ & 22 & 3 & 570310 & 1.65 \\
BLG326 & 17\uph41\upm29\zdot\ups1 & $-39\arcd32\arcm00\arcs$ & 350.5114 & $-4.8171$ & 20 & 4 & 570484 & 1.62 \\
BLG327 & 17\uph44\upm28\zdot\ups1 & $-39\arcd32\arcm00\arcs$ & 350.8120 & $-5.3083$ & 20 & 3 & 566849 & 1.51 \\
BLG328 & 17\uph47\upm27\zdot\ups1 & $-39\arcd32\arcm00\arcs$ & 351.1089 & $-5.8018$ & 20 & 3 & 556223 & 1.36 \\
BLG329 & 17\uph50\upm26\zdot\ups1 & $-39\arcd32\arcm00\arcs$ & 351.4022 & $-6.2976$ & 19 & 2 & 540815 & 1.32 \\
BLG330 & 17\uph28\upm00\zdot\ups1 & $-29\arcd30\arcm00\arcs$ & 357.4584 & $ 2.9392$ & 113 & 3 & 1489220 & 2.31 \\
BLG331 & 17\uph24\upm50\zdot\ups1 & $-29\arcd20\arcm00\arcs$ & 357.2127 & $ 3.6048$ & 110 & 15 & 1514420 & 2.12 \\
BLG332 & 17\uph10\upm00\zdot\ups5 & $-32\arcd59\arcm50\arcs$ & 352.3810 & $ 4.0909$ & 111 & 5 & 1015348 & 1.64 \\
BLG333 & 17\uph35\upm30\zdot\ups1 & $-27\arcd10\arcm01\arcs$ & 0.3176   & $ 2.8305$ & 359 & 6 & 1270794 & 2.61 \\
BLG334 & 17\uph41\upm40\zdot\ups2 & $-24\arcd59\arcm53\arcs$ & 2.8940   & $ 2.8132$ & 107 & 7 & 1338590 & 2.57 \\
BLG335 & 17\uph34\upm32\zdot\ups9 & $-22\arcd29\arcm56\arcs$ & 4.1485   & $ 5.5217$ & 115 & 7 & 1007985 & 1.84 \\
BLG336 & 17\uph37\upm02\zdot\ups0 & $-22\arcd29\arcm58\arcs$ & 4.4558   & $ 5.0361$ & 114 & 6 & 1040076 & 1.97 \\
BLG337 & 17\uph39\upm31\zdot\ups0 & $-22\arcd29\arcm58\arcs$ & 4.7608   & $ 4.5497$ & 116 & 13 & 1087002 & 2.04 \\
BLG338 & 17\uph42\upm00\zdot\ups0 & $-22\arcd29\arcm59\arcs$ & 5.0632   & $ 4.0618$ & 113 & 6 & 1219047 & 1.98 \\
BLG339 & 17\uph47\upm00\zdot\ups0 & $-22\arcd29\arcm59\arcs$ & 5.6653   & $ 3.0756$ & 412 & 3 & 1261835 & 1.98 \\
BLG340 & 17\uph49\upm30\zdot\ups2 & $-22\arcd29\arcm54\arcs$ & 5.9644   & $ 2.5806$ & 314 & 3 & 1185439 & 2.25 \\
BLG341 & 17\uph51\upm57\zdot\ups9 & $-22\arcd29\arcm59\arcs$ & 6.2539   & $ 2.0911$ & 109 & 5 & 1260449 & 2.38 \\
BLG342 & 17\uph46\upm29\zdot\ups7 & $-23\arcd05\arcm22\arcs$ & 5.0996   & $ 2.8703$ & 325 & 3 & 1372416 & 1.92 \\
BLG343 & 17\uph48\upm59\zdot\ups8 & $-23\arcd05\arcm21\arcs$ & 5.3969   & $ 2.3775$ & 318 & 3 & 1295253 & 2.15 \\
BLG344 & 17\uph43\upm30\zdot\ups3 & $-23\arcd40\arcm55\arcs$ & 4.2355   & $ 3.1480$ & 251 & 3 & 1503135 & 2.04 \\
BLG345 & 17\uph46\upm00\zdot\ups4 & $-23\arcd40\arcm56\arcs$ & 4.5339   & $ 2.6589$ & 109 & 2 & 1458394 & 2.06 \\
BLG346 & 17\uph42\upm00\zdot\ups3 & $-24\arcd16\arcm29\arcs$ & 3.5501   & $ 3.1290$ & 332 & 3 & 1457492 & 2.15 \\
BLG347 & 17\uph44\upm31\zdot\ups3 & $-24\arcd16\arcm25\arcs$ & 3.8517   & $ 2.6410$ & 314 & 3 & 1400744 & 2.23 \\
BLG348 & 17\uph44\upm30\zdot\ups2 & $-20\arcd04\arcm55\arcs$ & 7.4377   & $ 4.8242$ & 114 & 2 & 1001745 & 1.68 \\
BLG349 & 17\uph45\upm31\zdot\ups2 & $-19\arcd29\arcm30\arcs$ & 8.0689   & $ 4.9247$ & 108 & 2 & 896882 & 1.71 \\
BLG350 & 17\uph47\upm50\zdot\ups2 & $-21\arcd00\arcm00\arcs$ & 7.0531   & $ 3.6820$ & 109 & 2 & 1138004 & 1.91 \\
BLG351 & 17\uph43\upm20\zdot\ups2 & $-21\arcd29\arcm50\arcs$ & 6.0825   & $ 4.3217$ & 103 & 2 & 1166359 & 1.80 \\
BLG352 & 17\uph39\upm20\zdot\ups2 & $-21\arcd04\arcm56\arcs$ & 5.9472   & $ 5.3327$ & 183 & 2 & 1055393 & 1.64 \\
BLG353 & 17\uph49\upm30\zdot\ups2 & $-17\arcd39\arcm59\arcs$ & 10.1278  & $ 5.0490$ & 101 & 6 & 767562 & 1.59 \\
BLG354 & 17\uph35\upm01\zdot\ups4 & $-23\arcd34\arcm58\arcs$ & 3.2886   & $ 4.8483$ & 307 & 3 & 1132575 & 2.09 \\
BLG355 & 17\uph26\upm00\zdot\ups2 & $-25\arcd24\arcm56\arcs$ & 0.6228   & $ 5.5703$ & 107 & 10 & 1040564 & 1.86 \\
BLG356 & 17\uph25\upm00\zdot\ups1 & $-27\arcd44\arcm51\arcs$ & 358.5517 & $ 4.4605$ & 106 & 6 & 1181464 & 2.09 \\
BLG357 & 17\uph23\upm50\zdot\ups3 & $-25\arcd59\arcm52\arcs$ & 359.8641 & $ 5.6521$ & 105 & 3 & 1067907 & 1.69 \\
BLG358 & 17\uph18\upm59\zdot\ups8 & $-29\arcd00\arcm05\arcs$ & 356.7655 & $ 4.8440$ & 133 & 7 & 1297380 & 1.74 \\
BLG359 & 17\uph16\upm23\zdot\ups1 & $-29\arcd00\arcm00\arcs$ & 356.4383 & $ 5.3130$ & 21 & 5 & 1026221 & 1.65 \\
BLG360 & 17\uph13\upm45\zdot\ups5 & $-29\arcd00\arcm06\arcs$ & 356.1036 & $ 5.7808$ & 111 & 17 & 1010778 & 1.55 \\
BLG361 & 17\uph19\upm00\zdot\ups1 & $-29\arcd35\arcm30\arcs$ & 356.2797 & $ 4.5064$ & 19 & 7 & 988799 & 1.95 \\
BLG362 & 17\uph16\upm22\zdot\ups1 & $-29\arcd35\arcm30\arcs$ & 355.9504 & $ 4.9754$ & 20 & 6 & 1046220 & 1.79 \\
BLG363 & 17\uph13\upm43\zdot\ups5 & $-29\arcd35\arcm34\arcs$ & 355.6157 & $ 5.4436$ & 112 & 11 & 1131273 & 1.60 \\
BLG364 & 17\uph13\upm40\zdot\ups1 & $-32\arcd00\arcm03\arcs$ & 353.6439 & $ 4.0516$ & 112 & 13 & 1095239 & 1.80 \\
BLG365 & 17\uph09\upm30\zdot\ups0 & $-32\arcd00\arcm02\arcs$ & 353.1234 & $ 4.7669$ & 113 & 12 & 964653 & 1.58 \\
BLG366 & 17\uph06\upm29\zdot\ups9 & $-32\arcd44\arcm57\arcs$ & 352.1412 & $ 4.8308$ & 115 & 11 & 911721 & 1.51 \\
\hline}

The number of frames taken of each field should be considered with
caution. This is due to the observational strategy adopted for OGLE-III
phase which allowed to switch to a much denser, sometimes even nearly
continuous coverage of fields in which extremely interesting microlensing
events were detected online through our early warning (EWS and EEWS)
systems. For these fields, the total number of frames is not strictly
related to their typical observing frequency.

\vspace*{-3pt}
\Section{Photometric Maps of the Galactic Bulge Fields}
\vspace*{-3pt}
\Subsection{Catalog}
\vspace*{-3pt}
The construction of the general catalog -- photometric map of the Galactic
bulge was done similarly to the procedure applied to other main targets of
the OGLE-III observations, the Magellanic Clouds and the Galactic disk. The
procedure was detailed in Udalski \etal (2008b). Here, we only recall the
main features and describe small changes.

The catalog tables were created for every subfield (chip nos.\ 1--8)
separately. This is a logical consequence of the fact that images of each
mosaic chip were considered as individual parts of every main field as
listed in Table~1. The reference images, reductions and databases were also
made separately for every subfield.

As it was mentioned above, about 7\% of the total area covered in the
Galactic bulge is overlapped by more than one field. Accordingly, a few
percent of all objects can be found in more than one subfield catalog. The
current, text version of the photometric maps does not provide
cross-identification of those objects but we plan to include it in the
online version of the database which will be available in near future.

The catalog was constructed based on the {\it I}-band databases. Adding the
{\it V}-band photometry required cross-identification between objects in
{\it I} and {\it V} databases. The matching star in {\it V} was chosen to be the
closest object in $0\zdot\arcs5$ (1.9 pixel) radius. For objects identified
in more than one (overlapping) subfield we were able to construct single
{\it V} data sets of independent measurements for each star.

The data in the catalog tables include: star ID (column~1); equatorial
coordinates J2000.0 (2,3); $X,Y$ pixel coordinates in the {\it I}-band
reference image (4,5); photometry data {\it V}, $V-I$ and {\it I} (6,7,8),
{\it V}-band photometry parameters: number of points used for mean
magnitude (9), number of $5\sigma$-rejected points (10) and standard
deviation $\sigma$ (11); {\it I}-band photometry parameters, same as for
{\it V} (12, 13, 14). Values of 9.999 and 99.999 mark no data available in
the given column. Values prefixed with plus sign in column (10) indicate
multiple {\it V}-band cross-identification. Table~2 presents a sample of
catalog table for field BLG100.1.

The mean photometry was derived for all objects with minimum of 1 and 6
observations in the {\it V} and {\it I}-band, respectively, by averaging
all observations after removing $5\sigma$ deviating points. The limit for
{\it V} is so low because of the small overall number of {\it V}-band
observations in bulge fields. In the case of stars that had been
identified in {\it V}-band on more than one subfield, all {\it V}
observations formed a single data set and then were averaged because they
are independent measurements. Finally, the color term correction was
applied for each object to its database average magnitude according to the
transformation equations and color term coefficients presented in Udalski
\etal (2008a). The $(V-I)$ color derived from the database {\it I} and {\it
V} values was converted to the standard system using
$(V-I)=(\vv-i)/(1-\epsilon_V+\epsilon_I)$ formula. Then, {\it V} and {\it
I} magnitudes were adjusted by adding $\epsilon_{V,I}\cdot(V-I)$ term to
the database value. For objects that do not have color information the
average $(V-I)$ of the field was used for color correction of respective
{\it I} or {\it V} magnitude. For other OGLE-III targets (Magellanic
Clouds, Galactic disk) the mean value for the whole population was used but
for the Galactic bulge the differences in mean $(V-I)$ are so big (see
Table~1) that we decided to use proper mean values for each field. The
``database'' values mentioned here are the instrumental magnitudes
corrected for subtle effects discussed in detail by Udalski \etal (2008a).

The catalog includes all objects which are present in the {\it I}-band
databases. With the additional criteria for the quality of the photometry,
described above, a small number of objects (less than 2\%) entered the
catalog with no photometry given, marked by 99.999 values in both {\it I}-
and {\it V}-bands. For the sake of simplicity and to keep the star IDs
consistent with the databases, we have not removed these objects from the
catalog.

The astrometry was done in the same way as in previously published OGLE-III
Photometric Maps, using third-order transformation based on 2MASS Catalog
(Skrutskie \etal 2006). The details are given in Udalski \etal (2008a).

\begin{landscape}
\renewcommand{\arraystretch}{0.95}
\MakeTableSepp{
r@{\hspace{15pt}}
c@{\hspace{15pt}}
c@{\hspace{15pt}}
r@{\hspace{15pt}}
c@{\hspace{15pt}}
c@{\hspace{15pt}}
r@{\hspace{15pt}}
c@{\hspace{15pt}}
c@{\hspace{15pt}}
r@{\hspace{15pt}}
c@{\hspace{15pt}}
r@{\hspace{15pt}}
c@{\hspace{15pt}}
c}{12.5cm}
{OGLE-III Photometric Map of the BLG100.1 field (sample)}
{\hline
\noalign{\vskip4pt}
ID & RA    & DEC   & $X$~~~~ & $Y$ & {\it V} & $V-I$ & {\it I} & $N_V$ & 
$N^{\rm bad}_V$ & $\sigma_V$ & $N_I$ & $N^{\rm bad}_I$ & $\sigma_I$\\
   &(2000) & (2000)&&&&&&&&&&&\\
\noalign{\vskip4pt}
\hline
\noalign{\vskip4pt}
1  & 17\uph51\upm01\zdot\ups01 & $-30\arcd16\arcm16\zdot\arcs9$ & 389.87 & 17.04 & 19.583 & 5.371 & 14.212 & 6 & 0 & 0.049 & 1330 & 0 & 0.053 \\
2  & 17\uph51\upm01\zdot\ups20 & $-30\arcd15\arcm39\zdot\arcs4$ & 533.88 & 27.13 & 15.919 & 1.969 & 13.950 & 7 & 0 & 0.007 & 1547 & 4 & 0.015 \\
3  & 17\uph51\upm01\zdot\ups60 & $-30\arcd16\arcm17\zdot\arcs5$ & 387.42 & 46.62 & 13.684 & 0.824 & 12.860 & 6 & 1 & 0.002 & 1849 & 1 & 0.018 \\
4  & 17\uph51\upm01\zdot\ups73 & $-30\arcd16\arcm07\zdot\arcs6$ & 425.45 & 52.99 & 15.017 & 1.100 & 13.916 & 7 & 0 & 0.006 & 1915 & 0 & 0.014 \\
5  & 17\uph51\upm02\zdot\ups25 & $-30\arcd16\arcm15\zdot\arcs0$ & 396.82 & 78.91 & 99.999 & 9.999 & 14.807 & 0 & 0 & 9.999 & 1381 & 0 & 1.042 \\
6  & 17\uph51\upm02\zdot\ups39 & $-30\arcd17\arcm29\zdot\arcs8$ & 109.36 & 84.93 & 13.512 & 0.761 & 12.751 & 8 & 0 & 0.005 & 2147 & 0 & 0.017 \\
7  & 17\uph51\upm02\zdot\ups47 & $-30\arcd15\arcm45\zdot\arcs5$ & 510.54 & 89.89 & 14.610 & 1.132 & 13.479 & 8 & 0 & 0.002 & 2192 & 3 & 0.013 \\
8  & 17\uph51\upm03\zdot\ups11 & $-30\arcd16\arcm33\zdot\arcs5$ & 325.87 & 121.45 & 17.709 & 3.546 & 14.162 & 8 & 0 & 0.022 & 2307 & 0 & 0.013 \\
9  & 17\uph51\upm03\zdot\ups79 & $-30\arcd16\arcm57\zdot\arcs0$ & 235.10 & 154.61 & 17.066 & 3.359 & 13.708 & 8 & 0 & 0.007 & 2373 & 1 & 0.009 \\
10 & 17\uph51\upm03\zdot\ups93 & $-30\arcd17\arcm05\zdot\arcs3$ & 203.24 & 161.44 & 19.237 & 5.418 & 13.818 & 8 & 0 & 0.065 & 2380 & 0 & 0.081 \\
11 & 17\uph51\upm04\zdot\ups71 & $-30\arcd16\arcm18\zdot\arcs6$ & 382.82 & 201.19 & 19.248 & 4.728 & 14.520 & 8 & 0 & 0.048 & 2386 & 0 & 0.015 \\
12 & 17\uph51\upm05\zdot\ups13 & $-30\arcd16\arcm10\zdot\arcs9$ & 412.40 & 222.06 & 18.699 & 4.592 & 14.107 & 8 & 0 & 0.032 & 2385 & 1 & 0.009 \\
13 & 17\uph51\upm05\zdot\ups47 & $-30\arcd17\arcm19\zdot\arcs6$ & 148.21 & 237.78 & 14.309 & 1.732 & 12.577 & 8 & 0 & 0.006 & 2382 & 1 & 0.010 \\
14 & 17\uph51\upm05\zdot\ups98 & $-30\arcd15\arcm56\zdot\arcs2$ & 468.60 & 264.46 & 17.975 & 4.669 & 13.306 & 8 & 0 & 0.021 & 2387 & 0 & 0.043 \\
15 & 17\uph51\upm05\zdot\ups99 & $-30\arcd15\arcm37\zdot\arcs4$ & 541.07 & 265.11 & 14.553 & 1.294 & 13.259 & 8 & 0 & 0.002 & 2382 & 5 & 0.016 \\
16 & 17\uph51\upm06\zdot\ups13 & $-30\arcd15\arcm48\zdot\arcs3$ & 499.02 & 271.90 & 18.801 & 4.937 & 13.864 & 8 & 0 & 0.085 & 2387 & 0 & 0.038 \\
17 & 17\uph51\upm06\zdot\ups21 & $-30\arcd16\arcm13\zdot\arcs8$ & 401.11 & 275.78 & 14.912 & 0.840 & 14.072 & 7 & 1 & 0.003 & 2387 & 0 & 0.011 \\
18 & 17\uph51\upm06\zdot\ups34 & $-30\arcd15\arcm38\zdot\arcs4$ & 537.13 & 282.71 & 12.838 & 9.999 & 99.999 & 7 & 1 & 0.002 & 0 & 0 & 9.999 \\
19 & 17\uph51\upm07\zdot\ups04 & $-30\arcd16\arcm01\zdot\arcs1$ & 449.83 & 316.87 & 15.199 & 1.144 & 14.055 & 8 & 0 & 0.004 & 2386 & 1 & 0.012 \\
20 & 17\uph51\upm07\zdot\ups20 & $-30\arcd17\arcm08\zdot\arcs7$ & 189.55 & 323.89 & 17.192 & 3.975 & 13.216 & 8 & 0 & 0.069 & 2387 & 0 & 0.036 \\
21 & 17\uph51\upm07\zdot\ups41 & $-30\arcd17\arcm42\zdot\arcs2$ & 60.59 & 333.81 & 19.002 & 4.809 & 14.193 & 7 & 1 & 0.022 & 2026 & 2 & 0.022 \\
22 & 17\uph51\upm07\zdot\ups58 & $-30\arcd16\arcm48\zdot\arcs5$ & 267.43 & 343.34 & 14.695 & 0.929 & 13.766 & 7 & 1 & 0.003 & 2385 & 2 & 0.011 \\
23 & 17\uph51\upm08\zdot\ups22 & $-30\arcd17\arcm31\zdot\arcs7$ & 101.15 & 374.20 & 15.878 & 2.986 & 12.892 & 8 & 0 & 0.008 & 2334 & 10 & 0.007 \\
24 & 17\uph51\upm08\zdot\ups48 & $-30\arcd17\arcm10\zdot\arcs0$ & 184.32 & 387.71 & 16.824 & 3.591 & 13.233 & 8 & 0 & 0.009 & 2380 & 7 & 0.006 \\
25 & 17\uph51\upm08\zdot\ups46 & $-30\arcd15\arcm50\zdot\arcs3$ & 491.17 & 387.61 & 17.876 & 3.721 & 14.155 & 8 & 0 & 0.016 & 2381 & 6 & 0.008 \\
\noalign{\vskip4pt}
\hline}
\end{landscape}

\subsection{Improved Calibration of Red Objects}
As it was mentioned above and discussed in earlier papers (Udalski \etal
2002), the original calibration of the OGLE-II and OGLE-III photometry was
slightly inaccurate for very red or heavily reddened objects, especially in
the Galactic bulge where the interstellar extinction is large. This was
mainly due to the fact that the {\it I}-band filter used during the second
and third phases of the project was different from the standard
Kron-Cousins definition (Landolt 1992), exhibiting quite wide transparency
``wing'' into the infrared part of the spectrum. Furthermore, there are
practically no standard stars with $(V-I)>2$~mag because such red objects
are usually variable. Thus, the OGLE-II and OGLE-III {\it I}-band filter
photometry could not be calibrated precisely to the standard KC (Landolt)
system for stars redder than $(V-I)>2$~mag. Therefore extrapolated
transformation coefficients derived from the observed standard stars in the
$(V-I)<2$~mag range had to be used. This may, however, lead to the
systematic {\it I}-band deviations in the OGLE map {\it I}-band magnitudes,
as shown in Fig.~2 of Udalski \etal (2002).

Since the beginning of the OGLE-IV phase in 2010, the project
has been using the new, interference {\it V}- and {\it I}-band
filters. OGLE-IV {\it I}-band filter very closely reproduces {\it I}-band
definition adopted by Landoldt (1992) in his commonly used set of standard
stars -- the system to which the OGLE data are ultimately tied. We may
assume that the OGLE-IV {\it I}-band filter can be practically considered
the standard one.

When the first season of OGLE-IV observations was completed, we were able
to compare and recalibrate OGLE-III photometry. To this end, we
cross-identified a number of constant stars in selected OGLE-III and
OGLE-IV Galactic bulge fields. We chose several highly reddened fields
where the photometrically stable bulge red giants and disk main sequence
stars are shifted to high $(V-I)$ colors of 3--6. In this way we obtained
a huge set of photometric standards with extremely large $(V-I)$.

To double-check that OGLE-IV {\it I}-band filter can be indeed treated as
the standard one we compared OGLE-IV photometry with OGLE-III photometric
maps of selected LMC fields where the vast majority of stars are bluer than
$(V-I)=1.5$~mag. We found a constant shift of magnitudes of these two
datasets without any noticeable trend with the $(V-I)$ color indicating
that indeed OGLE-IV {\it I}-band filter approximates the standard one very
well.

To calibrate the OGLE-III photometry of redder objects we compared the mean
magnitudes from OGLE-IV photometry and OGLE-III uncalibrated maps of the
selected Galactic bulge fields. The mean magnitude difference at
$(V-I)=1.5$~mag was assumed as the zero point for a given field. In the
next step we merged the data from all compared fields and derived a mean
difference of OGLE-IV (standard) photometry and uncalibrated OGLE-III map
magnitudes and as a function of $(V-I)$ color. A second order polynomial
was fitted to this relation and used further as the calibration function
(Fig.~2):
$$\Delta I=-0.033918+0.016361(V-I)+0.004167(V-I)^2~{\rm [mag]}.$$

The correction value, $\Delta I$, reaches 0.02~mag at $(V-I)=2.15$~mag,
0.05 at 3.0, 0.1 at 4.0 and 0.2 at 6.0, respectively. We note that the
correction is approximately two times smaller than predicted in Udalski
\etal (2002).

Finally we used the derived relation for correction of {\it I}-band
photometry and $(V-I)$ color (for all objects with $(V-I)>1.5$~mag) of all
OGLE-III maps including the already released maps of the LMC, SMC and
Galactic disk.  The computed $\Delta I$ was added to {\it I} magnitude and
subtracted from $(V-I)$ color of every object in the catalog. We estimate
that the corrected {\it I}-band magnitudes are accurate to 0.01--0.02 for
$(V-I)<6$~mag.

\subsection{Summary and Data Presentation}
The OGLE-III Photometric Maps of the Galactic bulge fields contain entries
for about 340 million stars located in 267 OGLE-III fields. Figs.~3--5 show
the typical accuracy of the OGLE-III Photometric Maps of these targets:
standard deviation of magnitudes as a function of magnitude in the {\it V}-
and {\it I}-band for the fields BLG205.2, BLG245.7 and BLG292.5 of
relatively high, moderate and low stellar density, respectively.

The completeness of the photometry can be assessed from the histograms
presented in Figs.~6--8 for the same fields as in Figs.~3--5. It reaches
$I\approx 19$~mag and $V\approx 20.5$~mag. It is worth noting that a
significant number of objects (43\%) do not have {\it V} photometry. This
is mainly due to the large reddening of the Galactic bulge objects which
makes them undetectable on OGLE-III {\it V} frames. The small overall
number of {\it V} observations is another reason.

Figs.~9--11 present color--magnitude diagrams (CMDs) constructed for a few
selected subfields from different Galactic bulge fields observed by
OGLE-III survey. Additionally, Fig.~12 shows the CMD of one of the most
reddened OGLE-III fields, BLG179. It is located at $b=-1\zdot\arcd7$ where
the interstellar extinction starts rapidly increasing toward the Galactic
plane, eventually obscuring the whole Galactic bulge background. The red
giant branch and red clump in BLG179 field are, therefore, significantly
smeared out and elongated toward fainter magnitudes and redder $(V-I)$. To
emphasize the effect, this diagram is compiled from the data of all eight
subfields.

\Section{Data Availability}
The OGLE-III Photometric Maps of the Galactic bulge are available to the
astronomical community from the OGLE Internet Archive:

\begin{center}
{\it http://ogle.astrouw.edu.pl}\\
{\it ftp://ftp.astrouw.edu.pl/ogle3/maps/blg/}
\end{center}

The archives include the catalog tables with photometric data and
astrometry for each of the subfields. The {\it I}-band reference images are
also available. The usage of the data is fully allowed for the general
community. We only require the proper acknowledgment to the OGLE project.

We plan to build an online, interactive access to the photometric maps
database, allowing to retrieve objects fulfilling a user-defined set of
criteria. The availability of such an access will be announced on the OGLE
project WWW page.

\Acknow{The OGLE project has received funding from the European Research
Council under the European Community's Seventh Framework Programme
(FP7/2007-2013) / ERC grant agreement no. 246678 to AU.}

\vspace{8mm}
\noindent List of figures.

\vspace{2mm}
\FigCap{Map of OGLE-III Galactic bulge fields.}
\FigCap{{\it I}-band
magnitude correction for red objects. Large dots show the mean values of
differences taken from individual fields (small points). Solid line shows
the parabola fitted to the mean values. Dotted line marks zero correction
level for convenience.}
\FigCap{Standard deviation of magnitudes as a function of magnitude for a high density BLG205.2 field}
\FigCap{Standard deviation of magnitudes as a function of magnitude for a moderate density BLG245.7 field}
\FigCap{Standard deviation of magnitudes as a function of magnitude for
a low density BLG292.5 field}
\FigCap{Histogram of stellar magnitudes in the BLG205.2 field}
\FigCap{Histogram of stellar magnitudes in the BLG245.7 field}
\FigCap{Histogram of stellar magnitudes in the BLG292.5 field}
\FigCap{Color--magnitude diagram of the BLG205.2 field.}
\FigCap{Color--magnitude diagram of the BLG245.7 field.}
\FigCap{Color--magnitude diagram of the BLG292.5 field.}
\FigCap{Color--magnitude diagram of the highly reddened BLG179 field.}

\end{document}